\documentstyle[12pt]{article}
\begin{document}
\newcommand{\newc}{\newcommand}
 
\newc{\be}{\begin{equation}}
\newc{\ee}{\end{equation}}
\newc{\ba}{\begin{eqnarray}}
\newc{\ea}{\end{eqnarray}}
\newc{\bea}{\begin{eqnarray}}
\newc{\eea}{\end{eqnarray}}
\newc{\D}{\partial}
\newc{\ie}{{\it i.e.} }
\newc{\eg}{{\it e.g.} }
\newc{\etc}{{\it etc.} }
\newc{\etal}{{\it et al.} } 
\newc{\ra}{\rightarrow}
\newc{\lra}{\leftrightarrow}
\newc{\no}{Nielsen-Olesen }
\newc{\lsim}{\buildrel{<}\over{\sim}}
\newc{\gsim}{\buildrel{>}\over{\sim}}
 
\begin{titlepage}
\begin{center}
February 1998\hfill       
CRETE-98/13 \\
\hfill DEMO-HEP-98/05 \\
             \hfill    
hep-ph/9802395
\vskip 0.5cm
 
{\large \bf
Core Phase Transitions for Embedded Topological Defects}
\footnote{ Talk presented at the "Formation of Topological Defects" ESF
Network Meeting, Grenoble, France, September 1997.}

\vskip .1in
{\large Minos Axenides}\footnote{E-mail address:
 axenides@gr3801.nrcps.ariadne-t.gr },\\[.05in]
 
{\em Institute of Nuclear Physics,\\ N.C.R.P.S. Demokritos \\
153 10, Athens, Greece 
}

\vskip .1in
{\large Leandros Perivolaropoulos}\footnote{E-mail address:
leandros@physics.uch.gr},\\[.05in]
 
{\em Department of Physics\\
University of Crete\\
71003 Heraklion, Greece
}
\end{center}
\vskip .1in
\begin{abstract}
\noindent 
Vortices in superfluid 3He-B have been observed to undergo a core
transition. We discuss the analog phenomenon in relativistic field 
theories which admit {\it embedded} global domain walls, vortices 
and monopoles with a core phase structure.   
They are present in scalar field theories with 
approximate global symmetries which are broken 
both spontaneously and {\it in parts} 
explicitly.  For a particular range of parameters their
symmetric core exhibits an instability and decays into 
the nonsymmetric phase.
\end{abstract} 
\end{titlepage}

\section{Introduction and Conclusions}

In the superfluid phases of liquid $3He$ occur the most complicated known
vacuum states of condenced matter, in which many symmetries are 
simultaneously broken.These symmetries manifest in the physical 
properties of the quantized vortex lines in the two superfluid phases
of $3He$ under rotation. 

More specifically in the B-phase vortices
of two types appear: those with an axisymmetric core in the high 
pressure regime and the ones with a nonaxisymmetric core for low
pressure which possess distinct topological characteristics.
As such in passing from the high to the low pressure phase there
is a distinct core transition characterized by both a change in the 
rotational symmetry as well as of the topology of the core structure
\cite{vm82,sv87}. 
Direct experimental observation of such
a transition indicated the spontaneous breaking of axial symmetry
\cite{kk91}. The one quantum vortex (high pressure one) 
undergoes a dimerization into a pair of half-quantum vortices 
(low pressure vortices) producing a novel topological feature- 
the transformation
of point vortices or boojums into one another after circling either one
half quantum vortex. 

A first order phase transition usually follows
from such a change in the topology at the "bifurcation process" of the
vortex core. In equivalent terms this is thought of as a topological
transition between two inequivalent ways in which vorticity can flare
out into the momentum space \cite{vm82}.   

In this talk we present the analog phenomenon in quantum 
field theory. We review recent work of examples of global defects
which admit a core phase structure\cite{ap97,apt97}. It is a
result of deformations of the vacuum manifolds as a result of
partial explicit breaking of the global symmetries. More specifically global embedded defects such as domain walls, vortices
and monopoles  arise in scalar field theories 
that exhibit a partial
explicit breaking of an spontaneously broken 
global symmetry ,$ U(1)$ to $Z_2$ for domain walls 
$SU(2)$ to $U(1)$ for vortices and $SO(4)$ to $SO(3)$ for monopoles. 
For particular values of parameters
the defect core exhibits a transition, in analogy with their
superfluid $3He-B$ vortices, from a symmetric phase to a non-symmetric one. 
We will present our results in detail
for the case of global domain
walls vortices. 
In both cases we will identify the
parameter ranges for stability of the configurations with either
a symmetric or a non-symmetric core. 
For the case of a domain wall wall
we will discuss results of a simulation for an expanding bubble of a 
domain wall.

Interesting implications of such core phase transition for cosmic defects
relate to nontrivial dynamics between defects with nonsymmetric cores\cite{apt97}
as well as to novel realization of topological inflation\cite{l94,v94}
. Defects have been thought to seed an
inflating phase in the very early universe. This is because their
core is an effective trap of vacuum energy. In our case the ones which undergo
a core phase transition offer a novel kind of an inflating seed with
the core relaxing its vacuum energy and thus exiting from an otherwise eternally
inflating phase\cite{apt97}. Core reheating and the emergence of latent heat from a defect-core
are few of the interesting new phenomena that core phase transitions provides us.

\par

\section{Domain Walls with NonSymmetric Core}

We consider a model with a $U(1)$ symmetry explicitly broken to a $Z_2$. This
breaking can be realized by the Lagrangian density [\cite{vs94,ap97}]
\begin{equation}
{\cal {L}}={1\over 2}\partial _\mu \Phi ^{*}\partial ^\mu \Phi +
{=CC^2 \over 2}|\Phi |^2 + {m^2 \over 2}Re(\Phi ^2) - {h \over 4}|\Phi |^4
\end{equation}
where $\Phi =\Phi _1+i\Phi _2$ is a complex scalar field. After a rescaling
\begin{eqnarray}
\Phi &\rightarrow &{m\over \sqrt{h}} \Phi \\
x  & \rightarrow & {1\over m} x \\
M & \rightarrow & \alpha m
\end{eqnarray}

\par
The corresponding equation of motion for the field $\Phi$
is
\be
{\ddot \Phi} - {\nabla ^2} \Phi - (\alpha^2 \Phi + \Phi^*) +
|\Phi|^2 \Phi = 0
\ee

The potential takes the form
\be
V(\Phi) = -{m^4 \over {2
h}} (\alpha^2 |\Phi|^2 + Re(\Phi^2) - {1\over 2} |\Phi|^4)
\ee
For $\alpha
< 1$ it has the shape of a "saddle hat" potential
i.e. at $\Phi = 0$ there is a local
minimum in the $\Phi_2$ direction but a local maximum in the $\Phi_1$ (Fig
1). For this range of values of
$\alpha$ the equation of motion admits the well known  static kink solution
\bea
\Phi_1
&=& \Phi_R \equiv \pm (\alpha^2 + 1)^{1/2} \tanh (({{\alpha^2 + 1}\over 2})^{1/2}
x) \\
\Phi_2 &=& 0
\eea
It corresponds to a {\it symmetric} domain
wall since in the core of the soliton the full symmetry of the
Lagrangian is manifest ($\Phi(0) = 0$) and the
topological charge arises as a consequence of the behavior of the
field at infinity
($Q={1 \over 2}(\Phi(-\infty) - \Phi(+\infty))/ (\alpha^2 + 1)^{1/2} $).

For $\alpha > 1$ the local minimum in the $\Phi_2$ direction becomes a
local maximum but the vacuum manifold remains disconnected, and the $Z_2$
symmetry remains. This type of potential may be called a "Napoleon hat"
potential in
analogy to the Mexican hat potential that is obtained in the limit $\alpha
\rightarrow \infty$ and corresponds to the restoration of the $S^1$ vacuum
manifold.

\begin{figure}
\begin{center}
\unitlength1cm
\begin{picture}(6,3)
\put(-3.5,-7.0){\includegraphics{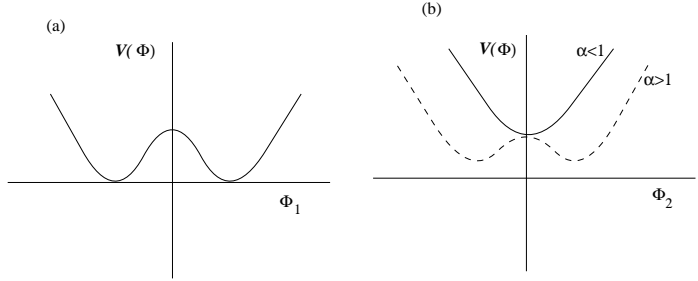}}
\end{picture}
\end{center}
\caption{(a) The domain wall potential has a local maximum
at $\Phi = 0$ in the $\Phi_1$ direction. (b) For $\alpha > 1$ ($\alpha 
< 1$)
this point
is a local maximum (minimum) in the $\Phi_2$ direction.
}
\end{figure}

The form of the
potential however implies that the symmetric wall solution may not be stable
for $\alpha >1$ since in that case the potential energy favors a solution
with $\Phi_2 \neq 0$. However, the answer is not obvious because for $\alpha
> 1$, $\Phi_2 \neq 0$ would save the wall some potential energy but would cost
additional gradient energy as $\Phi_2$ varies from a constant value at $x=0$
to 0 at infinity. Indeed a stability analysis was performed by
introducing a small
perturbation about the kink solution reveals the presence of negative modes
for $\alpha > \alpha_{crit}= \sqrt{3} \simeq 1.73$
 For the range of values $ 1 < \alpha < 1.73 $ the potential takes the shape
of a "High Napoleon hat".  
We study the full non-linear static
field equations obtained from (6) for a typical value of $\alpha=1.65$
  with boundary conditions
\bea
\Phi_1 (0)
&=& 0 \hspace{1cm} \lim_{x\rightarrow \infty} \Phi_1 (x) = (\alpha^2
+1)^{1/2} \\
{\Phi_2^\prime} (0) &=& 0 \hspace{1cm} \lim_{x\rightarrow
\infty} \Phi_2 (x) = 0
\eea

\begin{figure}
\begin{center}
\unitlength1cm
\begin{picture}(10,4)
\put(-3.0,-1.7){\includegraphics{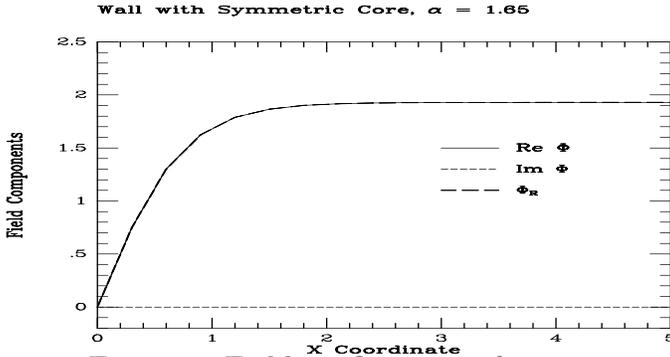}}
\end{picture}
\end{center}
\caption{Field configuration for a symmetric wall with $\alpha = 1.65$.
}
\end{figure}

Using a relaxation method based on collocation
at gaussian points [\cite{numrec}] to solve the system (6)
of second order non-linear
equations we find that for $ 1 <\alpha < \sqrt{3}$ the solution relaxes to the
expected form of (7) for $\Phi_1$ while $\Phi_2 = 0$ (Fig. 2). For $\alpha >
\sqrt{3}$ we find $\Phi_1 \neq 0$ and $\Phi_2 \neq 0$ (Fig. 3) obeying the
boundary conditions (13), (14) and giving the explicit solution for the
non-symmetric domain wall. In both cases we also plot the analytic soluti
on
(7) stable only for $\alpha < \sqrt{3}$ for comparison (bold dashed line)
. As
expected the numerical and analytic solutions are identical for $\alpha <
\sqrt{3}$ (Fig. 2).

\begin{figure}
\begin{center}
\unitlength1cm
\begin{picture}(10,4)
\put(-3.0,-1.7){\includegraphics{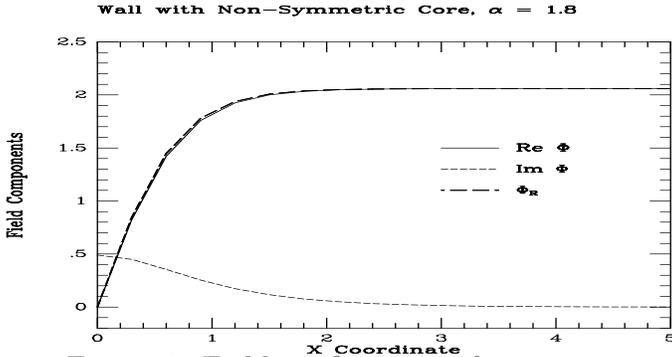}}
\end{picture}
\end{center}
\caption{Field configuration for a non-symmetric wall with $\alpha = 1.8$.
}
\end{figure}

\par
We now proceed to present results of our study on the
evolution of bubbles of a domain wall. We 
constructed a two dimensional simulation of the field evolution of domain
wall bubbles with both symmetric and non-symmetric core. In particular we
solved the non-static field equation (6) using a leapfrog algorithm
[\cite{numrec}]
with reflective boundary conditions. We used an $80 \times 80$ lattice
and in all
runs we retained ${{dt} \over {dx}} \simeq {1\over 3}$ thus satisfying the
Cauchy stability criterion for the timestep $dt$ and the lattice spacing $dx$%
. The initial conditions were those corresponding to a spherically symmetric
bubble with initial field ansatz
\be
\Phi (t_i) = (\alpha^2 + 1)^{1/2}
\tanh [({{\alpha^2 +1} \over 2})^{1/2} (\rho - \rho_0)] +
i \hspace{2mm} 0.1 \hspace{2mm}
e^{- ||x| - \rho_0|} {x \over {|x|}}
\ee
where $\rho = x^2 + y^2$ and $%
\rho_0$ is the initial radius of the bubble. Energy was conserved to within
2\% in all runs. For $\alpha$ in the region of symmetric core stability the
imaginary initial fluctuation of the field $\Phi (t_i)$ decreased and the
bubble collapsed due to tension in a spherically symmetric way as
expected.

\begin{figure}
\begin{center}
\unitlength1cm
\begin{picture}(6,8)
\put(-3.0,-2.0){\includegraphics{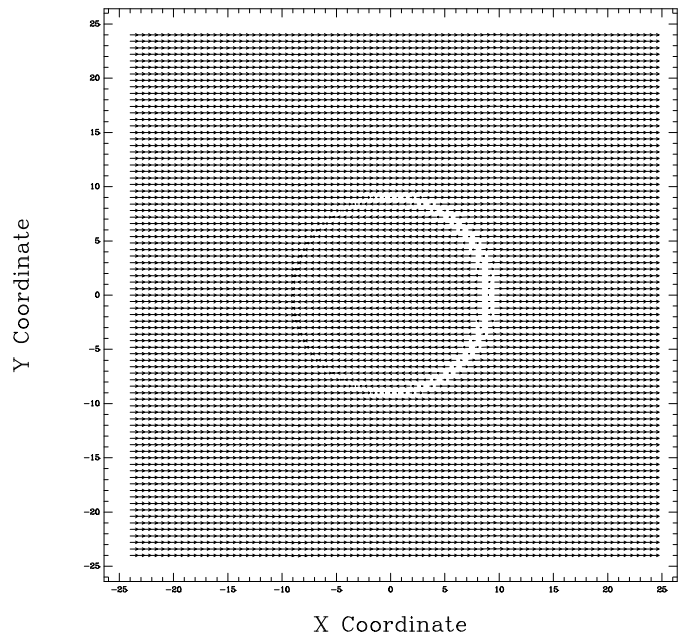}}
\end{picture}
\end{center}
\caption{Initial field configuration for a non-symmetric
spherical bubble wall with $\alpha = 3.5$.
}
\end{figure}

For $\alpha$ in the region of values corresponding
to having a non-symmetric stable core the
evolution of the bubble was quite different. The initial imaginary
perturbation increased but even though dynamics favored the increase of the
perturbation, topology forced the $Im\Phi (t)$ to stay at zero along a line
on the bubble: the intersections of the bubble wall with the y axis
(Figs. 4, 5). Thus in the region of these points, surface energy (tension)
of the bubble wall remained larger than the energy on other points of the
bubble. The result was a non-spherical collapse with the x-direction of the
bubble collapsing first (Fig. 5).

\begin{figure}
\begin{center}
\unitlength1cm
\begin{picture}(6,8)
\put(-3.0,-2.0){\includegraphics{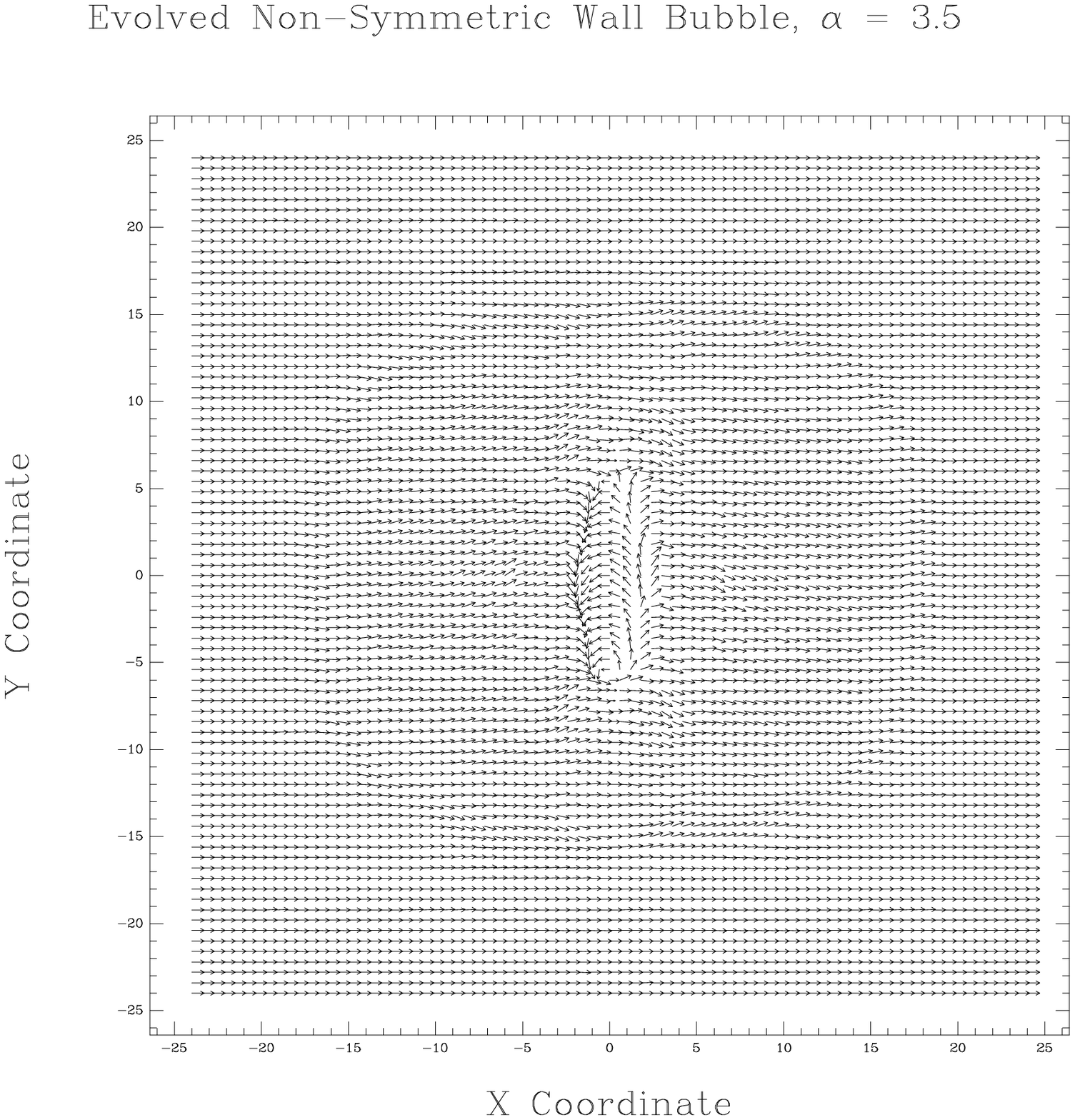}}
\end{picture}
\end{center}
\caption{Evolved field configuration ($t=14.25$, 90 timesteps) for a non-symmetric
initially spherical bubble wall with $\alpha = 3.5$.
}
\end{figure}

\section{Vortices with Nonsymmetric Core}

We have generalized our analysis for domain walls to the case
of a scalar field theory that admits global vortices. We consider a model
with an $%
SU(2)$ symmetry explicitly broken to $U(1)$. Such a theory is described by
the Lagrangian density:
\be
{\cal{L}} = {1 \over 2} \D_\mu {\Phi^\dagger}
\D^\mu \Phi + {M^2 \over 2} {\Phi^\dagger}\Phi + {m^2 \over 2} {\Phi^\dagger}
\tau_3 \Phi - {h\over 4} ({\Phi^\dagger}\Phi)^2
\ee
where $\Phi =(\Phi_1, \Phi_2)$ is a complex scalar doublet and $\tau_3$ is the $2 \times 2$
Pauli matrix. After rescaling as in
equations (2)-(4) we obtain the equations of motion for $\Phi_{1,2}$
vspace{2cm}
\begin{figure}
\begin{center}
\unitlength1cm
\begin{picture}(6,8)
\put(-3.0,-1.7){\includegraphics{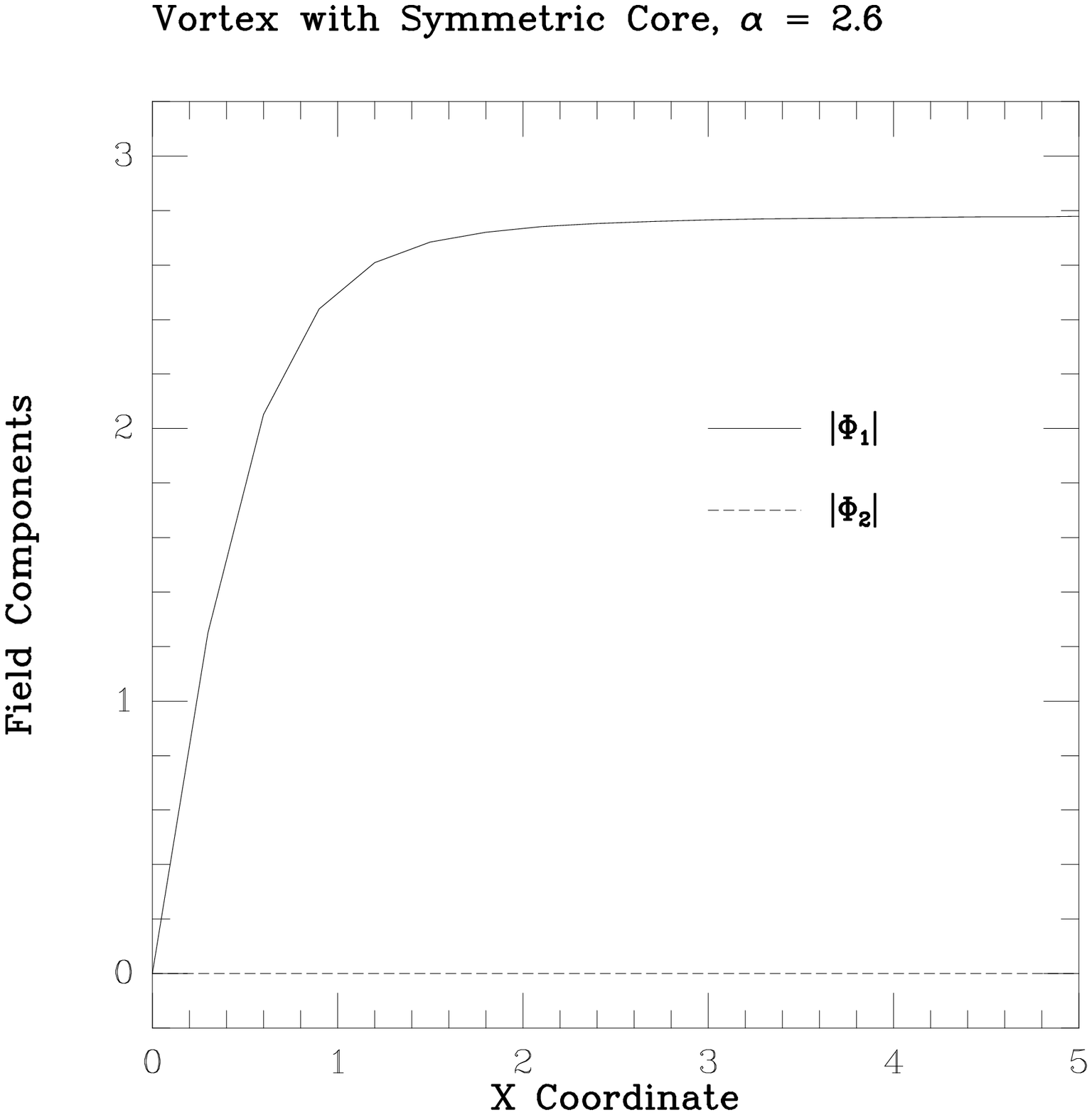}}
\end{picture}
\end{center}
\caption{Field configuration for a {\it symmetric-core}
global string with $\alpha = 2.6$.
}
\end{figure}

\be
\partial_\mu \partial^\mu \Phi_{1,2} - (\alpha^2 \pm 1) \Phi_{1,2}
+({\Phi^\dagger} \Phi) \Phi_{1,2} = 0
\ee
where the + (-) corresponds to
the field $\Phi_1$ ($\Phi_2$).
\par
Consider now the ansatz
\be
\Phi =
\left( \begin{array}{c}
\Phi_1 \\
\Phi_2
\end{array} \right)=
\left( \begin{array}{c}
f(\rho) e^{i\theta} \\
g(\rho)
\end{array}
\right)
\ee
with boundary conditions
\bea
\lim_{\rho \rightarrow 0} f(\rho)
&=& 0, \hspace{3cm} \lim_{\rho \rightarrow 0} { g^\prime} (\rho) = 0
\\
\lim_{\rho \rightarrow \infty} f(\rho) &=& (\alpha^2 +1)^{1/2},
\hspace{1cm} \lim_{\rho \rightarrow \infty} g (\rho) = 0
\eea

\par

This ansatz
corresponds to a global vortex configuration with a core that can be either
in the symmetric or in the non-symmetric phase of the theory. Whether the
core will be symmetric or non-symmetric is determined by the dynamics of the
field equations. As in the wall case the numerical solution
of the system ($21$) of non-linear complex field equations with the
ansatz $(22)$ for various values of the parameter $\alpha$
reveals the existence of an $\alpha_{cr} \simeq 2.7 $
For $\alpha < \alpha_{cr} \simeq 2.7$ the
solution relaxed to a lowest energy configuration with $g(\rho) = 0$
everywhere corresponding to a vortex with symmetric core (Fig. 6).

For $
\alpha > \alpha_{cr} \simeq 2.7$ the solution relaxed to a configuration
with $g(0) \neq 0$ indicating a vortex with non-symmetric core (Fig. 7). Both
configurations are dynamically and topologically stable and consist
additional paradigms of the defect classification discussed in the
introduction.

\begin{figure}
\begin{center}
\unitlength1cm
\begin{picture}(6,6)
\put(-3.0,-1.7){\includegraphics{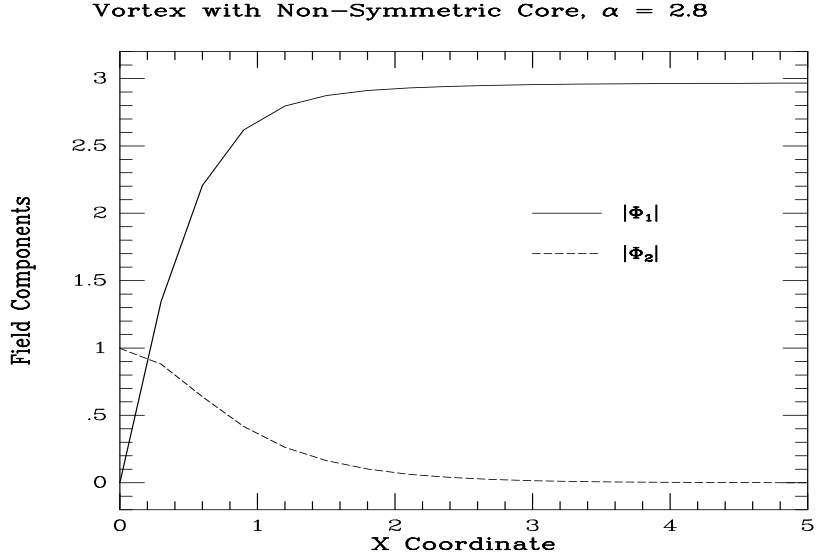}}
\end{picture}
\end{center}
\caption{Field configuration for a {\it non-symmetric-core}
global string with $\alpha = 2.8$.
}
\end{figure}
\par

\section{Acknowledgements}
This work was supported by the E.U. grants $CHRX-CT93-0340$, 
$CHRX-CT94-0621$. It is also a result of a network supported
by the European Science Foundation (ESF). The ESF acts as a
catalyst for the development of science by bringing together
leading scientists and funding agencies to debate, plan and
implement pan-European initiatives.

\end{document}